\documentclass[ aps, twocolumn, prd ,superscriptaddress]{revtex4}
\usepackage{mathptmx}
\usepackage{amsmath, amssymb, bm, graphicx, graphics, color,mathrsfs,hyperref,nicefrac}

\newcommand{\Lie}[0]{{\cal L}\, }

\newcommand{\tl}{\theta_{(\ell)}}
\newcommand{\tn}{\theta_{(n)}}

\newcommand{\nn}{\nonumber}
\newcommand{\be}{\begin{equation}}
\newcommand{\ee}{\end{equation}}
\newcommand{\bea}{\begin{eqnarray}}
\newcommand{\eea}{\end{eqnarray}}

\newcommand{\tq}{\tilde{q}}

\newcommand{\cV}{\mathcal{V}}

\begin{document}

\title{On the proximity of black hole horizons: lessons from Vaidya}

 
\date{\today}
\author{Ivan Booth} 
\email{ibooth@mun.ca}
\affiliation{Department of Mathematics and Statistics and Deparment of Physics and Physical Oceanography \\ Memorial University of Newfoundland,  
St.~John's, Newfoundland and Labrador, A1C 5S7, Canada}
\author{Jonathan Martin}
\email{jmartin@math.ualberta.ca}
\affiliation{Department of Mathematics and Statistics and Deparment of Physics and Physical Oceanography \\ Memorial University of Newfoundland, 
St.~John's, Newfoundland and Labrador, A1C 5S7, Canada}
\affiliation{
Department of Mathematical and Statistical Sciences, University of Alberta \\
Edmonton, Alberta, T6G 2G1, Canada}

\begin{abstract}

In dynamical spacetimes apparent and event horizons do not coincide. In this paper we propose a geometrical measure of the
distance between those horizons and investigate it for the case of the Vaidya spacetime. We show that it is well-defined, physically
meaningful, and has the expected behaviour in the near-equilibrium limit. We consider its implications for our understanding of
black hole physics. 

\end{abstract}


\maketitle

\section{Introduction}
If you pull on a rubber band, it stretches. 
As the Moon orbits Earth, its gravitational field raises the ocean tides. 
Gravitational waves impinging on an 
interferometer signal their presence by causing its arms to stretch and shrink. In each of these examples a physical object is distorted by
external forces, and that distortion signals a transfer of energy. 
Intuitively, one would expect black holes to behave in a similar way: tidal forces and gravitational waves should distort them and a distorted black 
hole should throw off gravitational waves as it rings down to equilibrium. Indeed numerical simulations and perturbative 
calculations show such behaviours (see, for example, \cite{NumEx} and \cite{PertEx} respectively). However at a fundamental level things are not so simple. 

\begin{figure}
\scalebox {.9}{\includegraphics{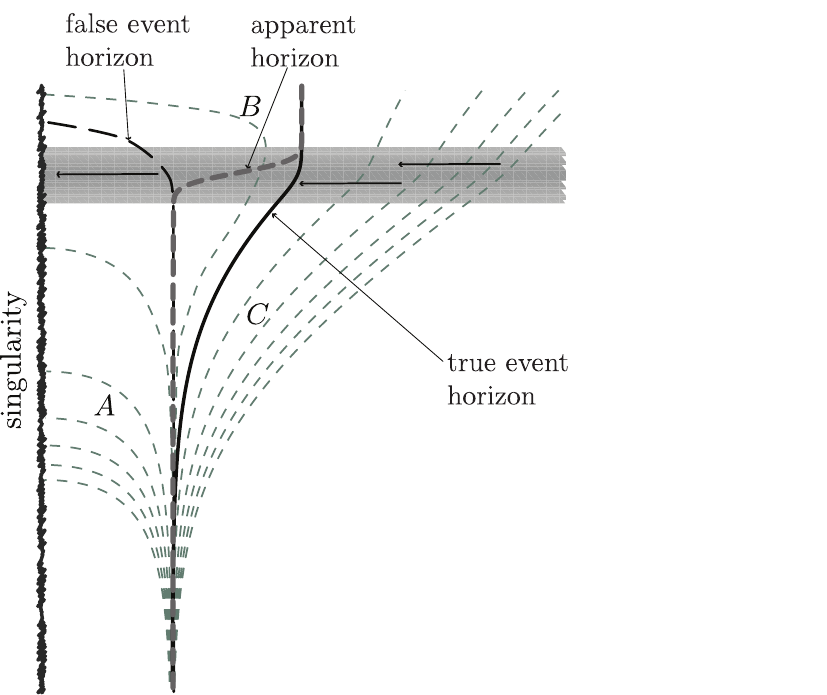}}\caption{A schematic showing the various horizons for a spherically symmetric 
spacetime with the angular dimensions suppressed. Horizontal location measures the areal radius of
the associated spherical shell while the direction of increasing time is roughly vertical outside the event horizon but 
tipping horizontal-and-to-the-left inside. The shaded gray region is a shell of infalling null dust, 
gray dashed lines represent ``outgoing'' radial null rays, and the apparent horizon is that associated with a spherically symmetric 
foliation of the spacetime. }
\label{EHFig}
\end{figure}

First, the physical objects discussed in the opening lines are 
locally defined: one can easily identify the edge of a rubber band or the surface of the Earth's ocean. This contrasts strongly with a classically 
defined black hole. 
A \emph{causal black hole} is a region of spacetime from which no (causal) signal can escape. Its boundary is the 
\emph{event horizon}. For outside observers, this null surface is the boundary between the unobservable events inside the black hole and 
those outside that can be seen. While very intuitive, a little thought shows that this definition is necessarily teleological: one determines
 the extent (or existence) of a black hole by tracing all causal paths ``until the end of time'' and then retroactively identifying any black hole region
(FIG.~\ref{EHFig}). Defined this way a black hole is a feature of 
the causal structure of the full spacetime rather than an independent object in its own right. 

The second complication follows from the first: a black hole does not directly interact with its environment. 
By definition it is causally disconnected from 
the rest of the spacetime and while it can be affected by outside events, signals from its interior cannot escape to influence the surrounding 
spacetime.
Again this contrasts quite strongly with our initial examples: one can see and feel a stretched rubber band and signals from the Earth can certainly
reach the Moon. 

The third complication also follows from the teleological definition of black holes: event horizons evolve in unexpected ways. As an example,   
consider FIG.~\ref{EHFig} again which depicts a simple spherically symmetric black hole which transitions between two equilibrium states by
absorbing a series of shells of infalling matter. Intuitively one might expect the event horizon to remain unchanging until the first matter arrives, grow
while it falls in, and then return to equilibrium when all of the matter has been absorbed. This is not what happens: the event horizon
is expanding before the matter arrives and its advent actually curtails, rather than causes, further expansion. 


In spite of  these apparent problems, it has already been noted that in practical calculations event horizons do seem to reflect physical 
interactions between a black hole and its surroundings in the way that one might expect. What then is happening in those examples? 
Well, as is recognized by the membrane paradigm \cite{membrane}, it is the external near-horizon gravitational fields 
(``just'' outside the event horizon) that are the agents of interaction with the environment. 
From this perspective the importance of horizons is not that they directly interact with the rest of the world, but rather that they might
closely reflect the physics of those near-horizon fields. For example, in an interaction with gravitational waves, a shearing of a horizon
should indicate a similar shearing of the gravitational fields ``just-outside'' the horizon. Further, a radial inflow of matter into a black hole
will increase the strength of those fields and result in a corresponding increase in horizon area. More topically, recent work has linked horizon
deformations to ``kicks'' in black hole mergers \cite{rezz}. 

Though this all seems plausible, we have already noted that, in general, event horizon evolution does not reflect local influences in such 
an obvious way. Understanding the regime where they will evolve in the expected way is one goal of this paper. In pursuit of this, we 
recall an alternate way to define black holes which is geometric and based on \emph{trapped surfaces}: closed and spacelike two-surfaces with
the property that all normal congruences of forward-in-time travelling null rays decrease in area. These surfaces are characteristic of black hole 
interiors. For example, a classic theorem by Penrose\cite{penrose} tell us that if the null energy condition holds, then those normal congruences
necessarily terminate at spacetime singularities. Further in asymptotically flat spacetimes, trapped surfaces are necessarily contained in causal
black holes\cite{wald, hawkellis} . 

Given a foliation of spacetime into ``instants'',  one can (in principle) locate all trapped surfaces at each instant and take their union 
to find the \emph{trapped region}. The boundary of that union is an \emph{apparent horizon} and can be shown to be \emph{marginally trapped} 
with vanishing outward null expansion\cite{wald,hawkellis}. Now in practical calculations, such as in numerical relativity, one cannot possibly 
locate all trapped surfaces. Instead one works with the outermost marginally trapped surface (which can be identified by 
well-known algorithms \cite{thornburg}). Even without an underlying spacetime foliation, one can study hypersurfaces foliated by marginally
trapped tubes as potential black hole boundaries. In general these are \emph{marginally trapped tubes} but specific types of them have been 
much studied over the last couple of decades including trapping\cite{hayward}, isolated\cite{isoref}, 
and dynamical horizons\cite{dynref}.


Now, these geometrically defined horizons can be locally identified and, just as importantly, they evolve in response to local stimuli. For example, 
in FIG.~\ref{EHFig} the geometric horizon expands if and only if matter is falling through it. This makes them attractive as candidates
for understanding black hole dynamics. However, there are still difficulties: 1) in asymptotically flat spacetimes on which the energy conditions hold,
marginally trapped surfaces are necessarily contained within event horizons and so still not in causal contact with the rest of spacetime and 
2) marginally trapped tubes are non-rigid and may be deformed \cite{gregabhay, bfbig,AHNU}. That is, unlike event horizons, they are non-unique.  

So from a theoretical perspective, both the causal and geometric definitions are problematic if we hope to understand black holes as astrophysical
objects interacting with their surroundings in intuitive ways. However, whatever theory says, in perturbative calculations, one sees both event 
\cite{PertEx} and geometric \cite{bill} horizons interacting with their surrounding in the way one might expect: shearing and expanding under the 
influence of gravitational waves. Perturbative calculations necessarily describe near-equilibrium black holes, and in this regime
the various theoretical problems somehow melt away. Indeed in the 
perturbative calculations, apparent and event horizons even coincide to a high degree of accuracy \cite{PertEx}. 

We then advocate the following model of near-equilibrium black holes. By definition, the near horizon fields of the membrane paradigm
are just outside the event horizon which in near-equilibrium should also be very ``close'' to the event horizon. 
In this regime we would expect that both the apparent and event horizons would closely reflect the evolution of the near-horizon fields and 
so be a useful tool for understanding black hole evolution. The goal of this paper is to (begin to) quantitatively understand the details of how 
this happens. In particular we will develop tools to geometrically quantify what ``close'' means. Though our ultimate goal is
to develop these ideas in general, in this paper we will focus on spherically symmetric spacetimes and in particular the Vaidya
solution as an initial testing ground. Some similar ideas have recently been advocated and developed in \cite{Nielsen} 
and will be discussed at appropriate points of this paper.

We proceed as follows. In Section II we review the mathematics of trapped surfaces and horizons, specializing to spherical symmetry 
which is all that we 
will need for this paper. In particular we consider the circumstances under which either event or apparent horizons may be considered to 
be near-equilibrium. Section III turns to Vaidya spacetimes, exactly locating their apparent horizons. In the appropriate regimes event horizons
are located either numerically or perturbatively. In both cases we again pay particular attention to the near-equilibrium regime and note that 
under those conditions the horizons are close as expected (at least in a coordinate sense). Section IV quantifies that observation by 
introducing a physically motivated, geometrically defined notion of the (proper time) distance between the horizons. We explore and refine this 
with numerical and perturbative calculations and prove that for near-equillibrium Vaidya spacetimes, the event and apparent horizons are 
indeed arbitrarily close together. Section V reviews the preceding sections 
and examines their further implications.

\section{Mathematics of spherically symmetric hypersurfaces}
\label{Geom}
As promised, we now review the mathematics of horizons: that is the geometry of two- and three-dimensional surfaces in a four-dimensional 
spacetime. Since this paper focuses on the Vaidya spacetime, we restrict our attention to spherically symmetric spacetimes and (hyper)surfaces
that share those symmetries. For details of derivations or the corresponding non-symmetric expressions see, for example, \cite{bfbig}. 

\subsection{Geometry of two-surfaces and three-surfaces}
\label{gg}

Let $(M,g_{ab})$ be a time-orientable four-dimensional spherically-symmetric spacetime.
Then given a spacelike two-sphere $S$ (which shares the spacetime symmetries), the normal space at each 
point has signature $(1+1)$ and can always be spanned by two future-oriented null vectors. We respectively label the
future-oriented outward and inward pointing null normals as $\ell$ and $n$, cross-normalize so that $\ell \cdot n = -1$ and require that 
they also share the spacetime symmetry: if $\zeta$ is any rotational Killing vector field then $\Lie_\zeta \ell = \Lie_\zeta n = 0$. 
There is a degree of freedom for these vectors: they will remain null, cross-normalized and with the correct symmetries  
under rescalings $\ell \rightarrow h \ell$  and $n \rightarrow n/h$ where $h$ is any function that shares the symmetries of $S$. 

The induced metric
on $S$ is that of a sphere and may be written in standard spherical coordinates as
\bea
\tilde{q}_{AB} = r^2 ([d\theta]_A \otimes [d\theta]_B + \sin^2 \theta [d \phi]_A \otimes [d\phi]_B ) 
\eea
where $r$ is the areal radius. The corresponding area element is 
\bea
\sqrt{\tilde{q}} = r^2 \sin \theta \, , 
\eea
and the (two-dimensional) Ricci scalar is
\bea
\tilde{R} = \frac{2}{r^2} \, . 
\eea

Next, consider the extrinsic geometry.  Then, the symmetry again significantly simplifies matters, and  
the only non-vanishing components of the extrinsic curvature are the expansions:
\begin{align}
&\tl = \frac{\Lie_\ell \sqrt{\tq} }{\sqrt{\tq}} = \frac{2 \Lie_\ell r}{r}   \; \mbox{and} \label{GenExp}
\\
&\tn = \frac{\Lie_n \sqrt{\tq} }{\sqrt{\tq}} = \frac{2 \Lie_n r}{r}  \, . \nonumber
\end{align}

Two-surfaces can be classified by the signs of these null expansions. 
If $n$ is inward pointing we expect $\tn < 0$ (areal radius gets smaller moving inwards) and classify the surfaces as:
\begin{enumerate}
\item \emph{untrapped} if $\tl>0$,
\item \emph{marginally trapped} if $\tl = 0$ and
\item \emph{trapped} if $\tl < 0$.  
\end{enumerate}
As mentioned in the introduction, the interiors of stationary black holes are made up of trapped surfaces while their boundaries may be 
foliated with marginally trapped surfaces.

Now consider a hypersurface $H$ foliated by spacelike two-spheres. Then for some function $C$ and an appropriate assigning of the null vectors,
we may always write a tangent vector to $H$  as 
\bea
\mathcal{V} = \ell^a - C n^a \, . 
\eea
The sign of $C$ determines the signature of the $H$: $C< 0 \Leftrightarrow$ timelike, 
$C= 0 \Leftrightarrow$ null and $C > 0 \Leftrightarrow$ spacelike. 
The rate of change of the area element ``up'' $H$ is 
\bea
\Lie_{\cV} \sqrt{\tq}  = \sqrt{\tq} \left( \tl - C \tn \right) \label{Lietq}
\eea
We will refer to $C$ as the \emph{evolution parameter}.

There are several other derivatives ``up'' the surface that are also of interest. First we have 
\bea
\kappa_{\cV} = -n_b \cV^a \nabla_a \ell^b \, .
\eea
As suggested by the notation, this reduces to the surface gravity (or equivalently the 
inaffinity of the scaling of the null vectors) for a Killing horizon.
Next the derivatives of the expansions are 
\begin{align}
\Lie_{\cV} \tl =& \kappa_{\cV} \tl  - \left( {\tl^2}/{2}   + G_{ab} \ell^a \ell^b \right)  \label{explderiv}
\\
&   + C \left( \frac{1}{r^2}  -  G_{ab} \ell^a n^b - \tl \tn \right) \nonumber \, , 
\end{align}
and 
\begin{align}
  \Lie_\cV \theta_{(n)}  = -  \kappa_\cV \tn + 
    C \left[  {\theta_{(n)}^{2} }/2 
    + G_{ab} n^{a} n^{b} 
     \right]  \label{expnderiv}  \\
  + \left[- \frac{1}{r^2} 
    + G_{ab} \ell^{a} n^{b} - \tl \tn \right] \,  . \nn  
\end{align}
These follow directly from Equations (2.23) and (2.24) of \cite{bfbig} (taking into account our spherical symmetry which 
sets shears, normal connections, and surface derivatives to zero). 
The $1/r^2$ comes from a two-dimensional Ricci scalar term.

\subsection{Spherically symmetric event horizons}

An event horizon is a null surface generated by null geodesics. In terms of the mathematics of the last section, it is a three-surface with $C=0$
so that 
\begin{align}
\Lie_{\cV} \tl =& \kappa_{\ell} \tl  - \left( {\tl^2}/{2}   + G_{ab} \ell^a \ell^b \right)   \label{explderiv2} \, . 
\end{align}
This is  the Raychaudhuri equation and it encapsulates the counter-intuitive behaviours of event horizons. Given the Einstein equations and
null energy condition and  adopting an affine scaling of the null vectors ($\kappa_\ell = 0$), it follows that 
\bea
\Lie_\ell \tl \leq - \frac{1}{2} \tl ^2  \label{explderiv3} \, . 
\eea
The rate of expansion of a null surface naturally decreases with time. There is nothing profound about this statement. The expansion as
defined by (\ref{GenExp}) is defined relative to the size of the sphere at the time of calculation. Thus the rate of expansion of a sphere 
expanding with constant speed naturally decreases, even in flat space. However, a non-zero flux of positive-energy density matter 
will cause the rate to decrease even more. Again this is to be expected: the gravitational influence of more mass inside the shell should 
decrease the rate of expansion. 

The second law also follows directly from Eq.~(\ref{explderiv3}). By this equation if $\tl$ is ever negative for a congruence of null curves, then that
congruence  will necessarily converge to a caustic. However, no event horizon can contain a caustic \cite{hawkellis}. Thus $\tl \geq 0$ and event 
horizons are non-decreasing in area. 

For a stationary event horizon $\tl = 0$ and so at first thought one might try to define a near-equilbrium horizon to be one with $\tl \ll 1/r$. However this
is, of course, not a geometrically meaningful statement since the null vectors retain a scaling freedom. Instead we return to (\ref{explderiv2}) and 
motivated by the thermodynamic analogy define:

{\bf Definition:} A spherically symmetric null surface is said to be \emph{slowly evolving} if
\bea
\Lie_{\ell} \tl \ll \kappa_\ell \tl  \; \; \mbox{and} \; \; \frac{1}{2} \tl^2 \ll  G_{ab} \ell^a \ell^b \, . \label{EHCond}
\eea

Note that this is a scaling-invariant statement: a quick check will show that if we rescale $\ell \rightarrow h \ell$ and $n \rightarrow n/h$, 
then the net effect is to rescale both conditions by a factor $h^2$ which will not affect the inequalities. 

If these conditions hold then, as first discussed in \cite{hawkhartle}, 
it is conventional to reinterpret Eq.~(\ref{explderiv2}) as a first law of event horizon mechanics with the 
(apparently) counter-intuitive behaviours only showing up at higher order. We find
\bea
\kappa_{\ell} \Lie_\ell \sqrt{\tq} =  \sqrt{\tq} \kappa_{\ell} \tl \approx \sqrt{\tq}  G_{ab} \ell^a \ell^b \, , \label{EH1}
\eea
which is of the same form as the $T \delta S = \delta E$ near-equilibrium version of the first law of thermodynamics\footnote{Alternatively, and perhaps
more logically, this should be interpreted as a Clausius-like statement of the second law: $\mathcal{L}_\ell \sqrt{\tilde{q}} \approx (\sqrt{\tilde{q}} G_{ab} \ell^a \ell^b)/\kappa_\ell$ in analogy with $\delta S \geq \delta Q / T$. See \cite{GourgTD} a discussion of this point of view.}. 
 Achieving this form is the main motivation for our definition. It will be further supported by later calculations but for now note that this first 
law is causal (to lowest order) with matter fluxes through the horizon driving the expansion. 

Next we consider geometrically defined horizons. 

\subsection{Spherically symmetric geometric horizons}

A \emph{marginally trapped tube (MTT)} is a three-surface that can be foliated with marginally trapped spacelike two-surfaces. There are several 
special cases of MTTs. Given the null energy condition, a null MTT is a \emph{non-expanding horizon} and physically corresponds to a black hole in
equilibrium with its surroundings \cite{isoref}. Dynamical black holes are represented by \emph{dynamical horizons} which are spacelike and 
expanding \cite{dynref}. \emph{Future outer trapping horizons (FOTHs)} can be either non-expanding
or dynamical and, in addition to $\tn < 0$, have $\Lie_n \tl < 0$ (there are fully trapped surfaces just inside the horizons) \cite{hayward}. 
Here we will
review some of their basic properties, but for a more general discussion see aforementioned original sources or
one of the review articles \cite{HorRev}. 

With $\tl = 0$, the expansion (\ref{Lietq}) of the MTT  is:
\bea 
\Lie_{\cV} \sqrt{\tq} = -\sqrt{\tq} C \tn  
\eea
and so with $\tn < 0$,
\begin{enumerate}
\item $C < 0 \Leftrightarrow$ timelike and contracting ($\Lie_{\mspace{-4mu} \cV} \sqrt{\tq} < 0$)
\item $C=0 \Leftrightarrow$ null and constant area  ($\Lie_{\mspace{-4mu} \cV} \sqrt{\tq} = 0$)
\item $C>0 \Leftrightarrow$ spacelike and expanding ($\Lie_{\mspace{-4mu} \cV} \sqrt{\tq} > 0$)  . 
\end{enumerate}
Further, since $\Lie_{\mspace{-4mu} \cV} \tl = 0$, $C$ can be found from (\ref{explderiv}) as 
\bea
C  =  \frac{G_{ab} \ell^a \ell^b}{{1}/{r^2} - G_{ab} \ell^a n^b} \, . \label{CEq}
\eea
For a marginally trapped surface 
\bea
\Lie_n \tl = - \left( \frac{1}{r^2} - G_{ab} \ell^a n^b \right) \, , \label{LNTL}
\eea
and so for a FOTH the denominator of Eq.~(\ref{CEq}) is positive by the assumption that there are fully trapped surfaces just inside the horizon. Violations
 of this condition only occur under extreme conditions such as the formation of new horizons outside of old ones \cite{mttpaper, AIPConf, harald}. 

Thus by (\ref{CEq}) we see that, if the null energy condition holds and $G_{ab} \ell^a \ell^b \neq 0$, a FOTH is necessarily a dynamical horizon 
(and vice
versa). On the other hand if $G_{ab} \ell^a \ell^b = 0$ then $C=0$ and the horizon is null, non-expanding and (weakly)
isolated. The causal nature of these geometric horizons is clear: they expand when matter falls in and are otherwise quiescent. Thus, like event
horizons, they cannot decrease in area and so obey a second law. 

Finally we consider the conditions under which a geometric horizon may be considered to be near-equilibrium. The conditions are
more complicated than for an event horizon but are well-studied \cite{SEH,bfbig, bill} and simplify greatly in spherical symmetry. 
Again we are guided by the requirement that a near-equilibrium horizon should obey a first law of the form (\ref{EH1}). To that end we combine
(\ref{explderiv}) and (\ref{expnderiv}) with $\tl =0$ to obtain
\bea
\kappa_{\cV} \theta_{(\cV)} - C \Lie_{\cV} \tn = G_{ab} \ell^a \ell^b - C^2 \left(  G_{ab} n^a n^b   +  \tn^2/2   \right) 
   \, . \label{kapT} 
\eea
Then we define a slowly evolving MTT as follows. 

{\bf Definition:} A spherically symmetric marginally trapped tube is said to be \emph{slowly evolving} if
\bea
 \Lie_{\cV} \tn \ll \left| \kappa_\cV \tn \right| \; \; \mbox{and} \; \;  C ( \tn^2/2 + G_{ab} n^a n^b ) \ll \frac{1}{r^2} \, . \label{CondII}
\eea

Again this is invariant under rescalings of the null vectors and with the help of (\ref{CEq}) is sufficient to reduce (\ref{kapT}) to a $T \delta S \approx \delta E$ form of the first law: 
\bea
\kappa_{\cV} \theta_{(\cV)} \approx G_{ab} \ell^a \ell^b \, . 
\eea

This first law is not the only motivation for this set of conditions. For example 
\bea
\frac{1}{2}  C \tn^2 \ll \frac{1}{r^2} \Longleftrightarrow \Lie_{\widehat{\cV}} r \ll 1 \, ,  
\eea
where $\widehat{\cV}^a$ is the unit-normalized version of $\cV^a$. That is, the rate of change of area radius is small relative
to distance measured ``up'' the horizon. Further discussion and motivation can be found in \cite{bfbig}. 

The non-uniquess of geometric horizons is not apparent from the above discussion. This is due to our spherical symmetry restrictions. For
spherically symmetric spacetimes there is a unique spherically symmetric marginally trapped tube. The other marginally trapped tubes don't share 
those symmetries. To allow our horizons to evolve into non-symmetric ones, we would need to
consider the general version of Eq.~(\ref{explderiv}) \cite{bfbig}.

With all of this mathematical background in mind, we turn to the Vaidya spacetime, which will be our chief example and testing ground for ideas. 

\section{Vaidya black holes and their horizons}
\label{vaidya}

In this section we identify and classify the horizons of the Vaidya spacetime.   
We start with a brief review of the spacetime itself along with some assumptions specific to our calculations.

\subsection{Vaidya spacetimes}


\begin{figure*}[t]
\scalebox {.85}{\includegraphics{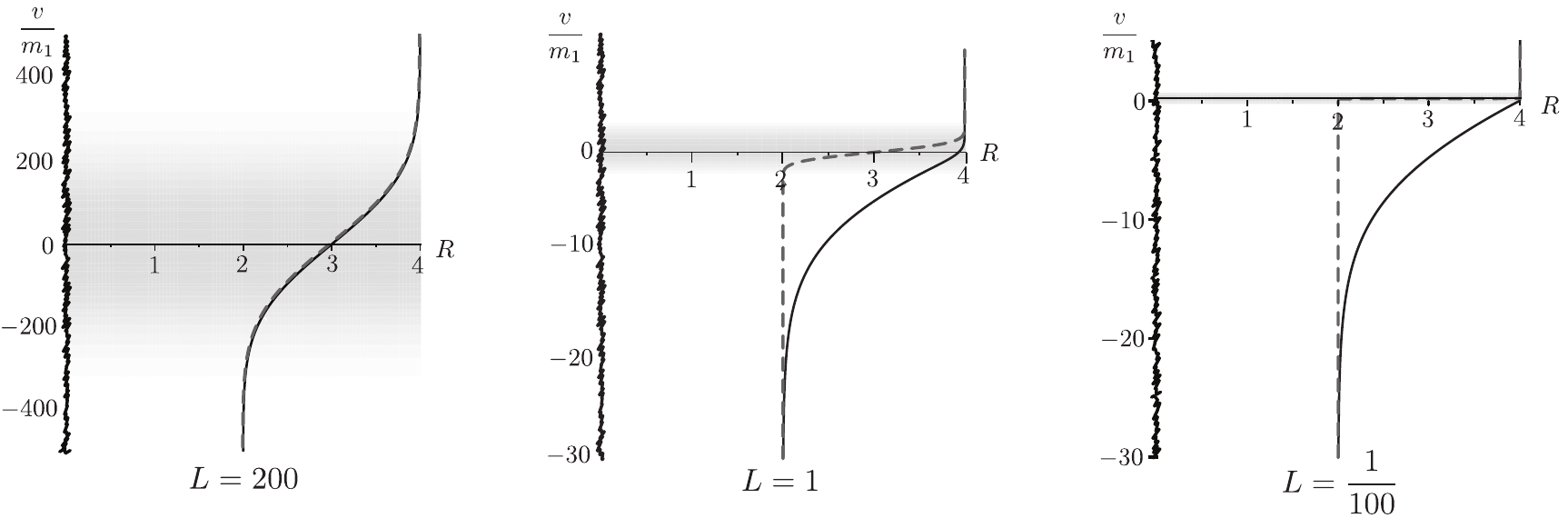}}
\caption{Event and apparent horizons for three transitions from $m_1$ to $m_2=2 m_1$. For ease of visual comparison (particularly 
for $L=1$ and $1/100$), the vertical axis is shown in terms of $v/m_1$ rather than $V$.
As in FIG.~\ref{EHFig}, the apparent horizon expansion 
is caused by the infalling matter, here represented by the shading (with density proportional to shading intensity).  
For small $L$ this also appears to be the case for event horizons. In contrast for large $L$, event horizon expansion is 
curtailed by the arrival of the matter shells.}
\label{HorizonsThree}
\end{figure*}

The Vaidya spacetime is described by the metric \cite{vaidya}
\bea
ds^2 &=& - \left(1- \frac{2m(v)}{r} \right) dv^2 + 2 dv dr + r^2 d \Omega^2  \label{metric}
\eea
and so has stress-energy tensor 
\bea
T_{ab} &=& \frac{dm/dv}{4 \pi r^2} [dv]_a [dv]_b \,  \label{stressenergy}
\eea
where $m(v)$ is a mass function and $d \Omega^2 = d \theta^2 + \sin^2 \theta d \phi^2$. 
Clearly if $m(v)$ is constant, this reduces to the Schwarzschild spacetime in ingoing Eddington-Finkelstein coordinates and as for 
that case,  $v$ is an advanced time coordinate labelling families of ingoing radial null geodesics. 
Applying the energy conditions to  (\ref{stressenergy}), one finds that the mass function must be non-decreasing. In this case, the matter content
of this spacetime consists of infalling shells of null dust.  

For most of this paper we will be interested in Vaidya black holes that transition from an initial mass $m_1$ to a final mass $m_2$. That is 
\bea
\lim_{v \rightarrow - \infty} m(v) = m_1  \; \; \mbox{and} \; \; \lim_{v \rightarrow \infty} m(v) = m_2 \, . 
\eea
We further assume that the evolution is characterized by a time-scale $\Lambda$ so that 
\bea
m(v) = m_1 M \left( \frac{v}{\Lambda} \right) \, , 
\eea
where we have adapted $m_1$ as a standard unit in order to write the mass function in a dimensionless form. In our 
study we will often also adopt dimensionless time 
\bea
V = \frac{v}{\Lambda} \, , 
\eea
which lets us to study a range of spacetimes for a given mass function $M(V)$, simply by changing the 
length scale $\Lambda$. In particular this allows for running to the near-equilibrium limit which we expect (and will see) to occur at
large $\Lambda$: in this case ``large'' means large relative to $m_1$. That is, setting 
\bea
\Lambda = m_1 L  \, , 
\eea
$\Lambda$ is large if $L \gg 1$. In later perturbative calculations we will use $(1/L)$ as the expansion parameter. 

It is convenient to continue this transformation into dimensionless quantities. Defining a new radial coordinate $R$ by 
\bea
r = R m_1  \, , 
\eea
the metric becomes
\bea
ds^2 = m_1^2 \left[ - \left(1 - \frac{2 M(V)}{R} \right) L^2  dV^2  + 2 L dV dR + R^2 d \Omega^2 \right] \, . \label{scaledmetric}
\eea
It turns out that having the scaling parameter $L$ appear directly in the metric in this way  (rather than inside the mass function) is computationally convenient. 

For the numerical calculations we will need a concrete form for the mass function. In that case we almost always assume that 
\bea
M(V) &=& \frac{3}{2} + \frac{1}{2} \mbox{erf} \left(  V \right)  \, , \label{massfunction}
\eea
where $\mbox{erf}$ is the error function. This describes a black hole irradiated with null dust that transitions from mass $m_1$ to mass $m_2 = 2 m_1$.

Next we locate and classify the horizons in Vaidya, beginning with the geometric horizons (which are easier). 

\subsection{Geometric horizons}
\label{GeoHorizons}

Let us consider the geometry of the surfaces of constant $v$ and $r$. First, 
a pair of (future-oriented) outward and inward pointing null normals are: 
\begin{align}
&{\ell} =  \frac{\partial}{\partial v}  +  \frac{1}{2}\left( 1 - \frac{2m(v)}{r} \right)   \frac{\partial}{\partial r} 
 \; \; \mbox{and } \label{NullVectors} \\
&n = - \frac{\partial}{\partial r}   \, .  \nonumber
\end{align}
These have the usual rescaling freedom of $\ell \rightarrow h \ell$ and $n \rightarrow n/h$ for any positive function $h(v,r)$, but the given forms 
are a computationally convenient choice. 

By (\ref{GenExp}) the corresponding expansions are
\begin{align}
\tl =  \frac{(r - 2 m)}{r^2}  \; \; \mbox{and} \; \;  \tn =  - \frac{2}{r} \,  \, . \label{tLtN}
\end{align}
Thus two-spheres are trapped if $r<2m$, untrapped if $r>2m$ and marginally trapped if $r=2m$. The marginally trapped surfaces foliate 
a marginally trapped tube for which Eq.~(\ref{LNTL}) gives 
\bea
\Lie_n \tl = -\frac{1}{4 m(v)^2} < 0 \, , 
\eea
(since $G_{ab} \ell^a n^b = 0$) and  as such it is a future outer trapping horizon. If $dm/dv = 0$ it is null, isolated, and non-expanding 
but when $dm/dv > 0$ it is spacelike, dynamical and expanding. These conclusions are confirmed by the evolution parameter which
from Eq.~(\ref{CEq}) is
\bea
C = 2 \frac{dm}{dv} \, \, . 
\eea
Several examples of these horizons are shown in FIG.~\ref{HorizonsThree}. As expected they expand in response 
to infalling matter and do not expand in the absence of such stimulus. 

We can also check the slowly evolving conditions to confirm that it is near-equilibrium for large $L$. For sufficiently large $L$ we have 
\begin{align}
& C \left( \frac{1}{2} \tn^2 + G_{ab} n^a n^b \right) = \frac{1}{L } \left( \frac{\dot{M}}{m_1^2 M^2} \right) \ll \frac{1}{4m_1^2 M^2} \; \; \mbox{and}\\
& \Lie_{\cV} \tn = \frac{1}{L} \left( \frac{\dot{M}}{m_1^2 M^2} \right) \ll   \frac{1}{4m_1^2 M^2} = \left| \kappa_{\cV} \theta_{(n)} \right| \, , \nn
\end{align}
where we have switched to dimensionless coordinates and a dot indicates a derivative with respect to $V$.
In this regime the FOTH is slowly evolving. 

\subsection{Event horizons}
\label{EH_Sect}

Next we locate event horizons in Vaidya. The key is to recall that the event horizon is 
necessarily a null surface and given the spherical symmetry of Vaidya, we expect this null surface to be similarly symmetric. From the metric,
spherically symmetric surfaces parameterized by $(v(\lambda), r(\lambda), \theta, \phi)$ are null if and only if they are tangent to one of 
the two null vector fields $\ell$ and $n$ defined in (\ref{NullVectors}).

The surfaces tangent to $n$ are clearly infalling with $v$=constant while the ``outgoing'' null surfaces are solutions of:
\bea
\frac{dr}{dv} = \frac{1}{2} \left(1 - \frac{2m(v)}{r} \right) \, . \label{OutNull}
\eea
Quotation marks are applied since for $r < 2m(v)$ these ``outgoing'' surfaces are also infalling (for example the curves labelled $A$ in FIG.~\ref{EHFig}).  This is consistent with our earlier observation that for $r< 2 m(v)$, $\tl < 0$. 

The event horizon must be an example of a surface tangent to $\ell$ since those tangent to $n$ all end up in the singuality. 
The trick is locating the correct null surface, and it is here that a knowledge of the future is important. Assuming that the black hole ultimately 
settles down (or at least 
asymptotes to) a Schwarzschild black hole of mass $m_2$, the event horizon should similarly settle down at $r = 2 m_2$. Thus, one takes 
$r(\infty) = 2 m_2$ as the (future) boundary condition. 

Generally this equation isn't solvable in exact form, however it can be solved both numerically for the mass function (\ref{massfunction}) and
then perturbatively in the near equilibrium regime. We do this in the next two subsections. 

\subsubsection{Numerical event horizons}

To solve (\ref{OutNull}) numerically we must first deal with the future boundary condition: $\lim_{v \rightarrow \infty} r(v) = 2 m_2$.
This can't be implemented exactly but it is sufficient to simply pick a large $v_o$ where $m \approx m_2$ and apply 
the boundary condition there. This works because, integrating backwards in time, null surfaces close to the event horizon
exponentially converge on it. 

This convergence property can be seen in FIG.~\ref{EHFig} however it can also be demonstrated by a perturbative calculation
(which in fact holds much more generally for perturbations about any given outgoing null geodesic). 
Consider a (spherically symmetric) null surface defined by 
$r(\lambda) = r_{EH} (\lambda) (1+ \xi(\lambda))$ where $r_{EH}(\lambda)$ is the true event horizon and $| \xi(\lambda) | \ll 1$ perturbs this to
a nearby null surface. Then by (\ref{OutNull}),
\bea
& & \frac{1}{\xi r_{EH}}\frac{d (\xi r_{EH}) }{dv} \approx \frac{m}{r_{EH}^2} \\
&& \Longrightarrow \xi \approx \xi_o  \left( \frac{r_o}{r_{EH}} \right) \exp \left(-\int^{v_o}_{v} \frac{m}{r_{EH}^2} dv \right)  \, , \nonumber
\eea
where $r_o = r_{EH}(v_o)$ and $\xi_o= \xi(v_o)$. 
Now, making our usual assumption that the black hole started out in an equilibrium state with $m=m_1 < m_2$, 
this can be (crudely) bounded as
\bea
\xi \lesssim \xi_o \left(\frac{m_2}{m_1}\right) \exp \left(\frac{m_1}{4 m_2^2} (v- v_o ) \right)\, ,
\eea
and so $\xi \rightarrow 0$ (into the past) with a timescale $\approx 4 m_2^2/m_1$. 
As such, one doesn't need to get the final boundary condition exactly right: one just needs to pick a $v$ for which $m_2 - m(v)$ is sufficiently small. 
In our examples $v = 10 \Lambda$ is more than sufficient: in that case $m_2 - m(10 \Lambda) \approx 10^{-45}$ and it asymptotes closer to the 
true event horizon with a ``time'' scale of $\Delta v = 16$. 

This kind of reverse integration is a standard way to find event horizons in numerical relativity \cite{thornburg}. 
With the boundary condition taken care of in this way, it is then straightforward to solve (\ref{OutNull}). Fourth-order Runge Kutta is sufficient and  
several examples of event horizons found in this way are shown in FIG.~\ref{HorizonsThree}. For large scaling parameter $L$, 
the MTT and event horizon are 
close together (at least in the coordinate sense). Further they both appear to evolve in the same way, expanding only in response to infalling 
matter. This is to be expected since this is a near-equilibrium regime. For small $L$ the horizons are further 
 apart and the presence of matter is seen to brake the expansion of the event horizon rather than drive it. Again this behaviour 
 is to be expected, since in this case the full Raychaudhuri equation governs the evolution.

\subsubsection{Slowly evolving event horizons}

We now consider the large $L$ limit. In terms of dimensionless quantities (\ref{OutNull}) becomes
\bea
\frac{1}{L} \left( \frac{dR}{dV} \right) = \frac{1}{2} \left(1 - \frac{2 M(V)}{R} \right) \, . \label{OutNullL}
\eea
This can be solved perturbatively in powers of $1/L$. To do this it is necessary to assume that over the range of interest, derivatives of 
$M(V)$ are at most comparable in size to $M(V)$ itself. 

For large $\Lambda$ we expect the event horizon to be close to the geometric horizon and so consider
solutions of the form 
\bea
R_{EH} = 2 M(V) \left(1 +  \frac{A(V)}{L} + \frac{B(V)}{L^2} + O \left(\frac{1}{L^3} \right) \right) \, , 
\eea
where we assume the coefficients (and their derivatives) are all much smaller than $L$. 
Then again representing derivatives of $M$ with respect to $V$ by dots, we can solve (\ref{OutNullL})  to find
\begin{align}
R_{EH} \approx  2M(V) \left( 1+  \frac{4 \dot{M}}{L} + \frac{32 \dot{M}^2 + 16 M \ddot{M} }{L^2}  \right)  \label{EHcandidate} \, ,
\end{align}
to second order. It is straightforward to extend this solution to higher orders, but for our purposes this is enough. 

At first glance there is something a bit strange about this solution: we started out solving for  general ``out-going'' null geodesic solutions 
but in the end came up with just a single solution. The loss of other solutions can be attributed to
the strict conditions imposed by the ansatz.
As we saw in the last section, radial null geodesics rapidly converge to the event horizon in the past but correspondingly rapidly diverge
into the future. Then it is our condition that derivatives of coefficients in the ansatz are comparable in size to the coefficients themselves
that eliminates all other candidates. We will return to this point in section \ref{GeodPert} where we perturbatively solve the general (radial) 
geodesic equation with a more general ansatz. 

That said, our sole solution is a good candidate for the event horizon: it is null to order $O(1/L^3)$ and has the correct asymptotic 
behaviour. It may also be familiar to the reader as the event horizon candidate obtained in \cite{MukundII} (and also for 
linear mass functions in \cite{Nielsen}). 
In \cite{MukundII} the out-going null curve equation (\ref{OutNull}) was solved using a derivative expansion that is essentially 
equivalent to our procedure. There it was noted that this is an event horizon that is defined by the local geometry of spacetime. This is true but
it is important to keep in mind that its identification as an event horizon depends on two assumptions: 1) at some point in the future 
$m(v) \rightarrow m_2$ and 2) the hierarchy of derivatives (or equivalently our ansatz) continues to hold into the future. Thus claiming that it is a 
locally defined event horizon is slightly misleading. Event horizons are always null surfaces and the property of being null is local. The problem with locating an event horizon is picking the correct null surface and it is this that requires knowledge of the future. This case is no different. 

Finally we check the slowly evolving conditions (\ref{EHCond}). Expanding with (\ref{EHcandidate}) and using some results 
from section \ref{GeoHorizons}, we find that to leading order in each quantity 
\begin{align}
&\tl \approx \frac{2}{L} \left(\frac{\dot{M}}{m_1 M} \right) \; \; , \; \; 
\kappa_\ell \approx \frac{1}{4m_1 M} \; \; , \; \; 
G_{ab} \ell^a \ell^b \approx \frac{1}{2L}  \left( \frac{\dot{M}}{m_1^2 M^2} \right) \; \; , \\
& \mbox{and} \; \; \Lie_\ell \tl \approx - \frac{6}{L^2} \left(\frac{\dot{M}}{m_1 M} \right)^2 \, , 
\nn 
\end{align} 
on the perturbative event horizon. Thus for large $\Lambda$, $\tl^2/2 \ll G_{ab} \ell^a \ell^b $ and 
$| \Lie_\ell \tl | \ll \kappa_\ell \tl$ and this horizon is also slowly evolving.

\section{Proper time between horizons}
\label{relate}

Finally we quantify the distance between the horizons and show that for large $L$ the horizons become arbitrarily close. 
We begin with some general theory about hypersurfaces and their normal geodesics.

\subsection{Mathematical background}
\label{bkgd}

A quick examination of FIG.~\ref{HorizonsThree} shows that for large $L$, the event and apparent horizons appear to be very close together 
while for small $L$ they are relatively far apart. This has a good geometric interpretation: the $r$-values measure the areal radii of the horizons, 
so comparing these at constant $v$ is actually comparing the areal radii of the horizons along the ingoing radial null geodesics. 
This is a possible way to compare horizons at least in spherical symmetry and has recently been considered in some detail in \cite{Nielsen,AHNU2}.
There it was shown that, for a range of matter models, in the near-equilibrium regime the two horizons will be close in area\footnote{There is a difference in emphasis between
\cite{Nielsen} and our paper. Here we are interested in the fractional difference between the horizons while \cite{Nielsen} is more interested in the
absolute difference and the implications that difference might have for entropy and thermodynamics in general. In absolute terms the difference 
may be quite large.}. 

\begin{figure}
\scalebox{0.9}{\includegraphics{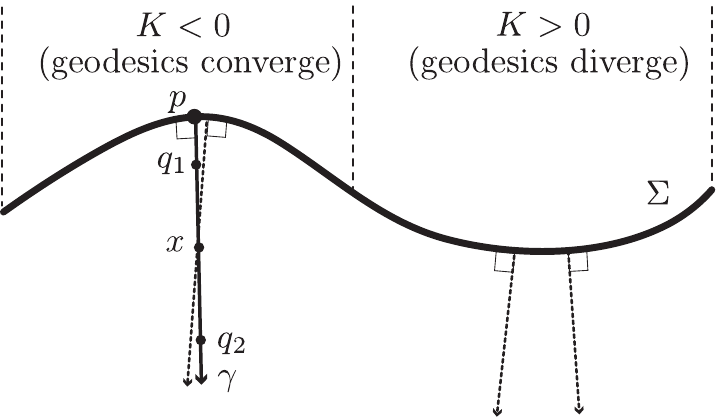}}
\caption{A spacelike surface $\Sigma$ 
in Minkowski space along with some of its timelike geodesic normals and a conjugate point. The normal vectors are
assumed to be past-oriented (as they are in our problem). By the theorem,  the maximum proper time from $q_1$ to $\Sigma$ is measured along 
$\gamma$. However, $q_2$ is beyond the conjugate point $x$ and so the maximum proper time from $q_2$ to $\Sigma$ may or may not be along 
$\gamma$. }  
\label{CPF}
\end{figure}

That said, this is not a measure of the distance between the horizons. Areal radius is a measure of area not distance and since curves of constant
$v$ are null, the geometric distance between any two points along such a curve is zero. For a distance we need a timelike or spacelike curve and 
in this case there is one obvious candidate: the timelike geodesic normal to the (spacelike) geometric horizon. With this candidate, the distance
between the horizons is the proper time measured (backwards in time) along this geodesic from the geometric to the event horizon. 
The use of this proper time as a measure of distance is supported by the following theorem\cite{wald}:

\textbf{Theorem 1:}  \emph{Let $\gamma$ be a smooth timelike curve connecting a point $q \in M$ to a point $p$ on a 
smooth spacelike hypersurface $\Sigma$. Then the 
necessary and sufficient condition that $\gamma$ locally maximize the proper time between $q$ and $\Sigma$ over smooth one-parameter
variations is that $\gamma$ be a geodesic orthogonal to $\Sigma$ with no conjugate point to $\Sigma$ between $\Sigma$ and $q$.} 

Recall that given the congruence of timelike geodesics normal to $\Sigma$ with unit tangent vector field $u^a$, a point $x$ on a particular 
geodesic $\gamma$ is \emph{conjugate} to $\Sigma$ if there exists an orthogonal (Jacobi) vector field $\chi^a$ along $\gamma$ that
 satisfies the geodesic deviation equation
\bea
u^a \nabla_a (u^b \nabla_b \chi^d) + \mathcal{R}_{abc}^{\phantom{abc}d} u^a \chi^b u^c = 0 \, ,  
\eea
and which vanishes at $x$ but not on $\Sigma$. $\mathcal{R}_{abc}^{\phantom{abc}d}$ is  the Riemann tensor. Intuitively these Jacobi 
fields generate the variations that map members of the congruence into each other, so a solution of these equations indicates the existence of 
two ``nearby'' members of the congruence that intersect at $x$. 

The theorem is demonstrated for the simple case of a Minkowski background in FIG.~\ref{CPF} which also demonstrates that 
it is a generalization of a Euclidean result: in (for example) three-dimensions the shortest distance between a point and a plane
is measured along the straight line through the point that is perpendicular to the plane. The switch to timelike geodesics means that 
the shortest distance becomes a longest one, and allowing for curved $\Sigma$ forces the inclusion of the  qualification re conjugate points. 

As suggested by that diagram, the sign of the (trace of the) extrinsic curvature can imply the existence of conjugate points. This is the \emph{focusing theorem} \cite{wald}:

\textbf{Theorem 2:} \emph{Let $(M,g_{ab})$ be a spacetime satisfying $\mathcal{R}_{ab} \xi^a \xi^b \geq 0$ for all timelike $\xi^a$ and 
let $\Sigma$ be a spacelike hypersurface with $K= \theta_{(u)} < 0$ at a point $p \in \Sigma$. 
Then within proper time $\tau \leq 3/|K|$ there exists a point $q$ conjugate to $\Sigma$ along the
geodesic $\gamma$ orthogonal to $\Sigma$ and passing through $p$, assuming that $\gamma$ can be extended that far. }

Paraphrasing, this theorem says that initially converging geodesics $K<0$ necessarily converge in finite time. In fact it
can be seen quite easily. For a hypersurface orthogonal congruence of timelike geodesics $u^a$,
the timelike Raychaudhuri equation tells us that
\bea
\frac{d \theta_{(u)}}{d \tau} = - \frac{1}{3} \theta_{(u)}^2 - \sigma_{(u)}^{\alpha \beta} \sigma^{(u)}_{\alpha \beta} - \mathcal{R}_{ab} u^a u^b \, .
\eea
In our case the geodesics start out being orthogonal to the spacelike horizon which has induced metric 
$q_{ij} = e_i^a e_j^b g_{ab}$ and extrinsic curvature 
$K_{ij} = e_{i}^a e_{j}^b \nabla_a u_b$  
($e_i^a$ is the pull-back operator onto the surface). Then at the point that they cross the horizon, the geodesics have 
expansion $\theta_{(u)} = K = q^{ij} K_{ij}$ and shear 
$\sigma^{(u)}_{ij} = K_{ij} - \frac{1}{3} K q_{ij}$. 
Thus if the dominant energy condition holds ($\mathcal{R}_{ab} u^a u^b \geq 0$), it is trivial that 
\bea
\frac{d \theta_u}{d \tau} \leq - \frac{1}{3} \theta_{(u)}^2
\eea
and the result follows by direct integration. 

\subsection{Finding the normal geodesics}
\label{FindBasic}
In order to apply the preceding theory, we first need to find the normal geodesics and it is to this problem that we now turn. 

Parameterizing with the non-affine (but otherwise 
very convenient) parameter $v$, the radial geodesic equations for the Vaidya spacetime are:
\bea
\frac{d^2 X^\alpha}{dv^2} + \Gamma^\alpha_{\beta \gamma} \frac{dX^\beta}{dv} \frac{d X^\gamma}{dv} = f \frac{dX^\alpha}{dv} \, ,  
\eea
where $X^\alpha(v) = [v, r(v),0,0]$ and on the right-hand side the extra function $f$ is required since $v$ is non-affine. From the first equation
($X^\alpha = v$) we immediately find that $f = \Gamma^v_{vv}$ and so general radial geodesics are solutions of
\bea
\frac{d^2 r}{dv^2} + (2 \Gamma^r_{vr} -  \Gamma^v_{vv}) \frac{dr}{dv} + \Gamma^r_{v v}  = 0 \, , \label{RadGeod}
\eea
where 
\bea
& & \Gamma^v_{vv} = \frac{m(v)}{r^2}  \; \; ,  \;  \;   \Gamma^r_{vr} = - \frac{m(v)}{r^2}  \mbox{and } \\
& & \Gamma^r_{v v} = \frac{\dot{m}(v)}{r} + \frac{m(v)}{r^3} \left(r - 2 m(v) \right)  \nonumber
\eea 
are the relevant non-vanishing Christoffel symbols. This is a general equation for radial geodesics and so solutions can be
timelike, spacelike or null. In particular the event horizon will be a null solution to these equations 
(in this case they are equivalent to Eq.~\ref{OutNull}). The parameterization by $v$ means that certain other solutions will be excluded: 
most notably the ingoing $v=$constant null geodesics. However, such solutions are not of interest to us in what follows.  

The timelike normal geodesics are solutions of these equations. Their initial conditions are determined by the timelike normal
to the spacelike geometric horizon. The future-oriented version of this normal is 
\bea
u =  \frac{1}{2 \sqrt{\dot{m}}} \left( \frac{\partial}{\partial v} \right)  - \sqrt{\dot{m}} \left( \frac{\partial}{\partial r} \right) \, ,
\label{u}
\eea
where here $\dot{m} = dm/dv$. Then the initial conditions for the timelike normal geodesics are
 \bea \left. r \right|_{v_f} = 2 m(v_f) \; \;  \mbox{and} \; \; \left. \frac{dr}{dv} \right|_{v_f} = - 2 \dot{m}(v_f)  \, . \eea

These equations are not solvable in closed form however it is straightforward to solve them numerically. We will be mainly interested in 
the mass function (\ref{massfunction}) and in particular studying how geodesic properties change with the scale parameter $L$. Then, 
as for the null geodesic equation, it is numerically and theoretically advantageous to switch to dimensionless form so 
that the mass function is invariant with respect to $L$ and the geodesic equations become
\bea
\frac{d^2 R}{dV^2} - \frac{3M L}{R^2} \frac{dR}{dV} + \left( \frac{L}{R} \frac{dM}{dV} + \frac{M L^2}{R^3} \left(R - 2M) \right) \right) = 0 \label{GeqII} \, , 
\eea
while the initial conditions maintain their original form
\bea \left. R \right|_{V_f} = 2 M(V_f) \; \;  \mbox{and} \; \; \left. \frac{dR}{dV} \right|_{V_f} = - 2 \left. \frac{dM}{dV} \right|_{V_f}   \label{ICII}
\, . \eea

For $L \lesssim 100$, a standard technique such as fourth-order Runge-Kutta is sufficient to solve the equations over much of the range of 
interest. However for larger values of $L$ or $| V | \gtrsim 5$ there are numerical difficulties. 
In both of those situations we enter a regime where all of the curves of interest are closer together than any reasonable numerical
accuracy that we might chose to work with. 
In such cases we turn to perturbative methods and consider the regime where  $R = 2 M(V) (1 + \rho(V))$ with $\rho(V) \ll 1$. The details of how
that expansion is done differs for the two cases and so we leave the details of this expansion to the appropriate sections.

\subsection{Spacetimes with $L \leq 100$}
\begin{figure}
\scalebox {.85}{\includegraphics{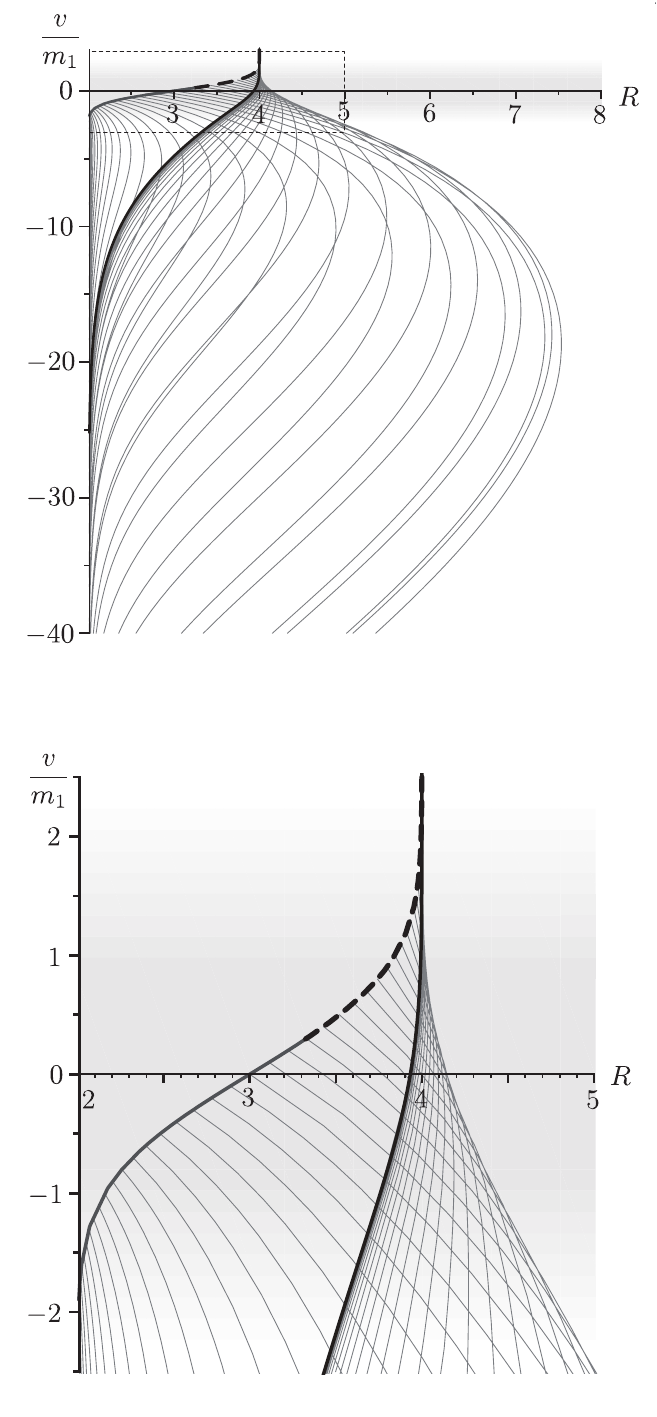}}
\caption{The geometric horizon and its timelike geodesic normals for $L=1$. 
The lower part of the figure is an enlargement of the boxed region in
the upper part of the figure. Nearby geodesics intersect outside the event horizon, suggesting the existence of conjugate points there. The dashed
part of the geometric horizon has $K_{(-u)}<0$ (past oriented-normal). 
  }
\label{ConjugatePoints}
\end{figure}
\subsubsection{Numerical regime}
We begin with the regime where the geodesic equations can be solved numerically. 
In particular we would like to know when (if) the normal
geodesics intersect the horizon and what is their proper time length; as set out in section \ref{bkgd} this will give a measure of distance between
the horizons. 

To study these issues, we numerically solve equation (\ref{OutNull}) with the 
appropriate (future) boundary condition to locate the event horizon. We then step along the geometric horizon repeatedly solving (\ref{GeqII}) 
with boundary conditions (\ref{ICII}) to find the normal timelike geodesics. Next, we search for intersections between the 
geodesics and event horizon. The Newton-Raphson method turns out to be sufficient for this purpose; if the numerics are reliable and
there is actually is an intersection, experience shows that this method quickly converges. With an intersection in hand, proper time length is easily 
found by integrating
\bea
T = \int_{v_{int}}^{v_f}  \left\{ \left( 1 - \frac{2m}{r}  \right) - \frac{dr}{dv} \right\}  dv\, , 
\eea
where $v_f$ is the value of $v$ at which the geodesic departs the geometric horizon and $v_{int}$ is the value where it intersects 
the event  horizon. If there are no conjugate points along these geodesics, then 
by Theorem 1 the time along any given geodesic will (at least locally) be the maximum possible proper time that can be measured along any
timelike curve connecting the intersection point to geometric horizon. 

\begin{figure*}
\scalebox {.85}{\includegraphics{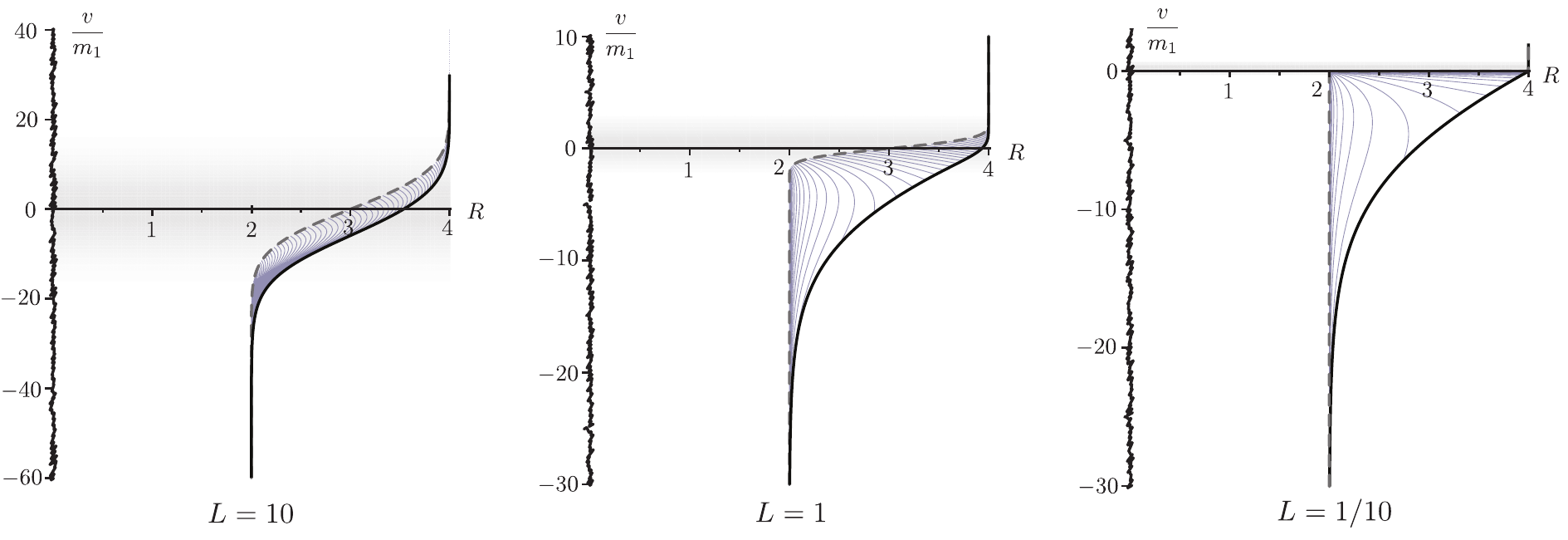}}
\caption{The timelike geodesics normal to geometric horizons for three representative values of the scale parameter $L$. Again the shading shows the concentration of infalling matter and the vertical axis is $v/m_1$ rather than $V$. To reduce clutter, the geodesics are cut off at the event horizon.} 
\label{HorizonsGeod}
\end{figure*}

This process is demonstrated in FIG.~\ref{ConjugatePoints} for the usual error-function-determined mass with $L=1$. The top half of the figure 
shows the normal geodesics over a wide $v$ range while the bottom half magnifies the region over which the geometric horizon is dynamically 
evolving. Keep in mind that this is a general coordinate system and so one should not jump to any conclusions based on the apparent shape
of curves. For example in this coordinate system the normal geodesics do not look like they are orthogonal to the geometric horizon (though of 
course they are) and the extrinsic curvature of the geometric horizon is not obvious. For future pointing normal $u^a$, the trace of the
extrinsic curvature is 
\bea
K_{(u)} = g^{ab} \nabla_a u_b = \frac{\dot{m} - 2 m \ddot{m} - 8 \dot{m}^2  }{8 m \dot{m}^{3/2}} \, ,
\eea
where in this case with $L=1$, dots are regular derivatives with respect to $v$.
The sign of this quantity is shown in the bottom half of the diagram and provides a nice demonstration of the focussing theorem. By that
theorem, conjugate points should exist along normal geodesics which exit the geometric horizon when $K_{(-u)} = - K_{(u)} < 0$ (the negative
sign is introduced since we are interested in past-oriented geodesics). A line of of these conjugate points can clearly be seen in the bottom
half of the figure and indeed these do occur on the expected geodesics, though outside of the event horizon. For the region over which 
$K_{(-u)}>0$, where the theorem makes no predictions, there do not appear to be any conjugate points at all. Thus in this example,
independent of the curvature of the apparent horizon, the normal geodesics maximize the proper time between the geometric and event horizons.

All other values of $L$ that we examined also lacked conjugate points between the horizons and so the geodesics in those cases continue to maximize proper time between the horizons (though see section \ref{double} for a more complicated situation). Three representative examples are shown 
in FIG.~\ref{HorizonsGeod}. Other trends apparent in those figures also continue for other values of $L$: for $L<1/10$ the apparent horizon
transition becomes sharper but otherwise the figure is more-or-less unchanged while for $L>10$ the curves become
closer together on the figure and the positive/negative $v$ behaviour becomes more symmetrical. 

It FIG.~\ref{HorizonsGeod} another important trend is clear: as $v_f$ decreases, the geodesics increasingly asymptote towards the event horizon
rather than clearly intersecting it. This effect can also be seen in FIG.~\ref{ConjugatePoints} where for larger values of $v_f$ the geodesics cross
the event horizon and then appear to be asymptoting ``back" towards it from the outside, while for more negative values they instead appear to be
asymptoting from the inside and it is not at all obvious that they ever intersect. The non-existence of an intersection in such a case is, of course,
very difficult to show numerically: however far we track the curves backwards in time (and so to whatever numerical accuracy) it is always 
possible that an intersection occurs just a bit further back (and so beyond any accuracy that we choose). Thus to demonstrate the existence
of such non-intersecting geodesics we turn to perturbative calculations. 

\subsubsection{Perturbative regime}
\label{GeodPert}

For negative $V$, we can asymptotically expand the mass function (\ref{massfunction}) and its derivatives  as
\begin{align}
M & \approx 1 + \frac{1}{2 \sqrt{\pi} V} \exp (-V^2) \, , \\
\frac{dM}{dV} &\approx \frac{1}{\sqrt{\pi}} \exp (-V^2)  \; \mbox{and}  \nonumber\\
\frac{d^2M}{dV^2} & \approx - \frac{2}{\sqrt{\pi}} V \exp (-V^2)  \nonumber \, . 
\end{align}
Then, for example, if $V = - 10$, $M-1$ and its derivatives are all less than $10^{-40}$. Meanwhile, as even a cursory 
examination of FIG.~\ref{ConjugatePoints} and FIG.~\ref{HorizonsGeod} demonstrates, if $R_{TG}$ is the location of the timelike geodesic then 
$R_{TG} - 2M$ will generally be much larger. In such a regime it is reasonable to take $M=1$ and set its derivatives to zero; 
essentially we revert to Schwarzschild geodesics. 

Then we look for perturbative solutions, expanding around $R=2$ (that is $r = 2m_1$):
\bea
R_{TG} = 2 \left(1 + \rho(V) \right) \, .
\eea
To first order in $\rho$ the geodesic equation (\ref{GeqII}) becomes
\begin{align}
& \frac{1}{L^2} \frac{d^2 \rho}{dV^2} - \frac{3}{4L}  \frac{d\rho}{dV} + \frac{\rho}{8}  \approx 0 \, , \label{GEpert1}
\end{align}
which has general solution
\bea
\rho \approx A_o \exp \left( \frac{LV}{4} \right) + B_o  \exp \left( \frac{LV}{2} \right) \label{rhoAp} \, , 
\eea
for some constants $A_o$ and $B_o$. Going back to the metric, it is easy to see that the value of $B_o$ determines the signature of the geodesic: 
$B_o > 0 \Leftrightarrow$ spacelike, $B_o = 0 \Leftrightarrow$ null, and 
$B_o < 0 \Leftrightarrow$ timelike. 

Next, in many of the numerically ambiguous cases, the regimes where the numerical and perturbative analyses work overlap and so we can use
these perturbative solutions to study the ultimate fate of the geodesics. For a $V_o$ in this overlap (typically $-10 < V_o < -40$ is used in our work)
we can determine the coefficients in the perturbative solutions by matching to the numerics. Doing this
\begin{align}
\rho^{EH}  = & \rho_o^{EH} \exp \left(\frac{L}{4} (V-V_o)\right)  \label{rhoEH} \; \mbox{and} \\
\rho^{TG} = & ( 2 \rho_o^{TG} - 4 \dot{\rho}_o^{TG}/L ) \exp  \left(\frac{L}{4}(V - V_o) \right) \label{rhoTG} \\
&  - ( \rho_o^{TG} - 4 \dot{\rho}_1^{TG}/L ) \exp  \left(\frac{L}{2} (V - V_o)\right) \nonumber \, ,
\end{align}
where $\rho_o = \rho(V_o)$ and $\dot{\rho}_o = \left. d \rho / dV  \right|_{V_o}$. 

Now, if no intersection has occurred for $V> V_o$, we can use these matched approximate solutions to study potential $V<V_o$ intersections. 
First note that using (\ref{rhoEH}) we can rewrite (\ref{rhoTG}) as
\begin{align}
\rho^{TG} =&  \left(  \frac{2 \rho_o^{TG} - 4 \dot{\rho}_o^{TG}/L}{\rho_o^{EH}} \right) \rho^{EH} \\
 &- ( \rho_o^{TG} - 4 \dot{\rho}_o^{TG} /L) \exp  \left(\frac{L}{2} (V - V_o)\right) \nonumber \,  .
\end{align}
By the discussion following (\ref{rhoAp}), $\rho_o^{TG} - 4 \dot{\rho}_o^{TG}>0$ and so we see that if a timelike geodesic arrives at $V_o$ 
with
\bea
\frac{2 \rho_o^{TG} - 4 \dot{\rho}_o^{TG}/L}{\rho_o^{EH}} < 1 \, , \label{cond}
\eea
and our approximation holds, then it will never intersect the event horizon. Numerical checks show that for $L=1$ this condition is satisfied 
for all geodesics which intersect the geometric horizon at $V_f \lesssim -1.087$ and not satisfied for larger values of $V_f$. Similar results
are found for the other values of $L$ and so this suggests the following picture: in each case there will be a transition point which separates
normal geodesics which intersect the event horizon from those that don't. 

In this situation one might guess that in the approach to the transition point from above, the intersection point diverges with
$V_{int} \rightarrow - \infty$. The perturbative solutions support this hypothesis. Solving (\ref{rhoTG}) for 
$\rho^{TG} = \rho^{EH}$ gives
\bea
\exp \left(\frac{L}{4} (V_{int}-V_o) \right) \approx \frac{2 \rho_o^{TG} - 4  \dot{\rho}_o^{TG}/L - \rho^{EH} }{\rho_o^{TG}- 4 \dot{\rho}^{TG}/L   
} \, . 
\eea
The denominator is always positive and the numerator vanishes exactly where the bound in condition (\ref{cond}) is saturated. Thus, as expected,
$V_{int} \rightarrow - \infty$ as $V_f$ approaches the transition point. 

\subsubsection{Proper time lengths}

\begin{figure}
\includegraphics{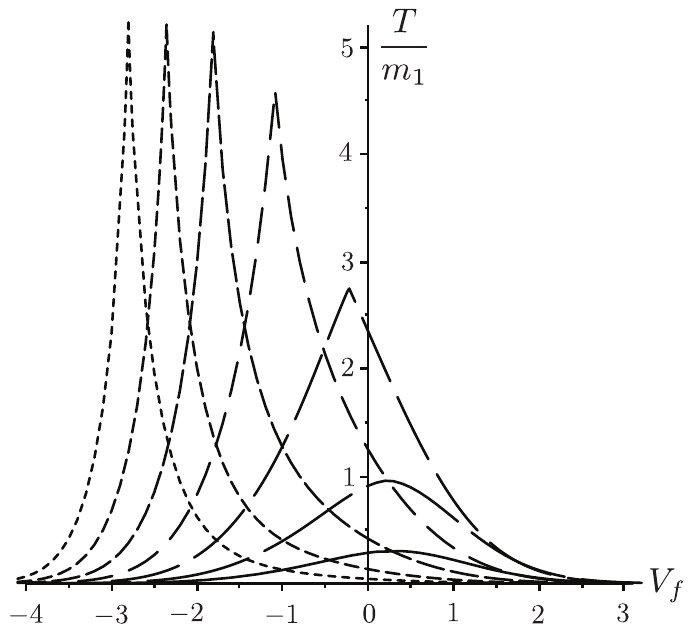}
\caption{Proper time lengths of normal geodesics versus departure point from the geometric horizon 
for seven values of the scale parameter $L$. Starting from the 
left-most and sharpest peak and ending with the right-most and flattest peak they are $L=10^{-3}$, $L=10^{-2}$, $L=10^{-1}$, $L=1$, $L=10^1$ and $L=10^2$, $L=10^3$. For $V > V_{peak}$ geodesics intersect the event horizon and their length between the horizons is show. For 
$V < V_{peak}$ there is no intersection with the event horizon. 
Instead the total length of the geodesic from $V=-\infty$ to $V_f$ is plotted.}
\label{Lengths}
\end{figure}
With this picture in mind we turn back to the numerics and results that were obtained for intersections and geodesic lengths. These results
are summarized in FIG.~\ref{Lengths} which requires a little explanation. First, for $L \leq 10$ there are sharp peaks with discontinuous first 
derivatives (there is also one for $L=100$, though it cannot be seen well in this figure). 
Those peaks are the demarcation between geodesics that ultimately intersect the event horizon (on the right) and those 
that don't (on the left). Computationally, they correspond to the point where the intersection finder first fails. However, within numerical error
they also match the transition points predicted above. To the right of the peak the length from $V_{int}$ to $V_f$ is plotted however on the
left it is instead the total length of the geodesic from $V=-\infty$ to $V_f$ (see Appendix \ref{appA} for a comment on how this is done). 
Intuitively one can think of these lengths as arising from a competition
between how close a curve is to being null versus its coordinate length. Thus for $V_f < V_{peak}$ all curves have infinite coordinate length
but are much closer to being null as $V_f$ decreases (have another look at the geodesics in FIG.~\ref{ConjugatePoints}). Meanwhile as $V_f$ 
increases past $V_{peak}$ the coordinate length of curves decreases and they ultimately become ``more null'' and so their decrease in length is
not unexpected. 

From the point of view of defining a maximum distance between the event and geometric horizons it is the $T(V_{peak})$ that is important.
Assuming that there are never any conjugate points between the horizons (we have checked many specific cases but not found a general proof),
then Theorem 1 tells us that from a given point on the event horizon, the (at least locally) maximum proper time from that point to geometric 
horizon is along a normal geodesic to the geometric horizon. This is exactly what is plotted in FIG.~\ref{Lengths}, though by the intersection 
with the geometric, rather than event horizon. Since the peak corresponds to $V=-\infty$ on the event horizon, it is only the part of the
curve $V_f > V_{peak}$ that is relevant to this discussion. Then $T(V_{peak})$ is the maximum proper time distance between the horizons
and this distance is measured along an ingoing timelike geodesic that asymptotes to the event horizon at $V=-\infty$ and ultimately orthogonally intersects the geometric horizon at $V_{peak}$. 

The second important observation to make from FIG.~\ref{Lengths} is that as $L$ increases, that maximum proper time decreases. This is slightly
qualified by the fact that numerical difficulties arise for $L=1000$ as the coordinate values
of the curves become extremely close together; in fact for $L=1000$ the algorithm doesn't find any intersections and so all integrals
are estimated using the methods of Appendix \ref{appA}. That said, the trend appears clear and this gives support to the hypothesis that 
for large enough $L$ the horizons become arbitrarily close. We will close the gap in this argument in the next section where we will employ 
perturbation techniques to study this large $L$ limit. 

\subsection{Large $L$ spacetimes}
\label{slow}

We now consider a regime where the entire geodesics can be described perturbatively.  In particular we confirm the conjecture that in the 
slowly evolving limit (as $L \rightarrow \infty$) the maximum distance between the geometric and event horizons goes to zero. 

\subsubsection{Geodesics}

To this end we return to 
the geodesic equation (\ref{GeqII}). This time we assume $\rho \sim \frac{1}{L} \ll 1$ and expand to third order (in each coefficient of 
the differential equation) as:
\begin{align}
& \frac{1}{L^2} \frac{d^2 \rho}{dV^2} + \left(-\underbrace{\frac{3}{4 M L}}_{O(\frac{1}{L})} + \underbrace{\frac{3 \rho}{2ML} + \frac{2 \dot{M} }{ML^2} }_{O(\frac{1}{L^2})}  - \underbrace{\frac{9 \rho^2}{4ML}}_{O(\frac{1}{L^3})} \right) \frac{d \rho}{dV} \nonumber \\
&+ \left(\underbrace{\frac{1}{8M^2}}_{O(1)} - \underbrace{\frac{3 \rho}{8M^2} + \frac{\dot{M}}{2M^2L} }_{O(\frac{1}{L})} 
+ \underbrace{\frac{3 \rho^2}{4M^2} - \frac{ \dot{M} \rho}{2M^2L} 
+ \frac{\ddot{M}}{ML^2} }_{O(\frac{1}{L^2})} \right) \rho \nonumber \\
& \approx   \underbrace{\frac{\dot{M}}{2M^2L} }_{O(\frac{1}{L})} - \underbrace{\frac{\ddot{M}}{ML^2} }_{O(\frac{1}{L^2})} \, ,
\label{GEpertII}
\end{align}
where $\dot{M} = dM/dV$ and $\ddot{M} = d^2M /dV^2$. 

To obtain a general solution of the geodesic equation we have to allow derivatives of $\rho$ to be much larger than $\rho$ itself. 
We have already seen this kind of behaviour during our first venture into perturbative solutions (\ref{rhoAp}). Inspired by that work but 
realizing that $M(V)$ is no longer constant, we begin by trying a first order solution of the form
\bea
\rho_{1} = \frac{4 \dot{M} }{L}  + \frac{1}{L} \left(f_{1/4}(V)  X + f_{1/2}(V) X^2  \right)  \label{rho1}
\eea
where $f_{1/4}(V)$ and $f_{1/2}(V)$ are functions of $V$ and all of the $L$ dependence is contained in the  
\bea
X=  \exp{\left( \frac{L}{4} \int_{V_f}^V \frac{d\nu}{M(\nu)} \right)} \, 
\eea
(this reduces to (\ref{rhoAp}) for $M=1$).
For normal geodesics the ``initial'' value in the integral will usually be the departure point from geometric horizon and so it is written as $V_f$.  
Substituting, it is straightforward to see that (\ref{rho1}) in its general form is a first order solution to (\ref{GEpertII}). This can be understood by 
noting that to lowest order, the $f_{1/4}$ and $f_{1/2}$ will be effectively constant relative to $X$ (for which changes in $V$ are magnified 
by the $L$).   

This freedom in the first order solution is eliminated by going to second order in the differential equation. Then, noticing the $\rho^2$ terms in (\ref{GEpertII}), we try a solution of the form:
\begin{align}
\rho_2 = & \rho_1 + \frac{1}{L^2} \left(32 \dot{M}^2 + 16 M \ddot{M}  \right) \label{rho2} \\ 
& + \frac{1}{L^2} \left(g_{1/4}(V)  X + g_{1/2}(V) X^2  + g_{3/4}(V)  X^3 + g_{1}(V) X^4 \right) \nonumber \, . 
\end{align}
This is a second order solution if and only if:
\begin{align}
&f_{1/4} = \frac{A_o}{M^3} \; \;  , \; \;  f_{1/2} = \frac{B_o}{M^4} \; \; , \; \;  g_{3/4} = - \frac{3 A_o B_o}{M^7}   \\
&\mbox{and} \;  \; g_1 = - \frac{B_o^2}{M^8} \nonumber \, , 
\end{align}
for arbitrary constants $A_o$ and $B_o$ and functions $g_{1/4}(V)$ and $g_{1/2}(V)$. 

Just as we had to go to second order to finalize the first order solution, we go to third order to finalize the second order solution. Then
\begin{align}
g_{1/4} = & \frac{1}{M^3} \left(\alpha_o - 4A_o \int_{V_f}^V \left\{ \frac{2 M \ddot{M} + \dot{M}^2}{M}  \right\} dv \right) \\
g_{1/2} = & \frac{1}{M^4} \left(\beta_o  +  2 \int_{V_f}^V \left\{ \frac{A_o^2 \dot{M}}{M^3} - 2 B_o \left(3 \ddot{M} + \frac{2\dot{M}^2}{M} \right)  \right\} \right) \nonumber
\end{align}
where $\alpha_o$ and $\beta_o$ are new constants and we omit the information on third order quantities since they won't be required in 
what follows.

This expression for general geodesics is quite formidable however we can extract useful information from it. First note that 
\bea
\exp \left(\frac{(V-V_f)L}{4(m_2/m_1)} \right) < X < \exp \left( \frac{(V-V_f)L}{4} \right) \, ,  
\eea
and so with $L$ large, all of these solutions will rapidly asymptote to $\rho_{EH}$ for $V < V_f$. Further they will rapidly diverge (and leave
the regime of validity of the approximation) for $V > V_f$. This helps to explain why $\rho_{EH}$ contains no free constants: it is the only 
solution that we have found that stays close to the geometric horizon. Other geodesics, whether timelike or null, diverge on a time scale of $1/L$. 

Now we focus our attention on the timelike normals. The initial conditions (\ref{ICII}) become 
\bea
\rho(V_F) = 0 \; \; \mbox{and} \; \; \left. \frac{d \rho}{dV} \right|_{V_f} = - \frac{2 \dot{M}_f}{M_f} 
\eea
where $M_f = M(V_f)$ and $\dot{M}_f = \dot{M}(V_f)$. Applying these we find that
\begin{align}
A_o = 0 \; \; \mbox{and}  \;\;  B_o = - 4 \dot{M}_f M_f^4 \; , 
\end{align}
while
\begin{align}
\alpha_o = -16 M_f^3 (\ddot{m}_f M_f- 2 \dot{M}_f^2) \;  \; \mbox{and} \; \; \beta_o = 16 \dot{M}_f^2 M_f^4 \, . 
\end{align}
Then, the timelike normal geodesic that intersects the geometric horizon at $V_f$ is parameterized (to second order in $1/L$) as
\begin{align}
\rho_{TG} = & \rho_{EH}  - \left(\frac{4 \dot{M}_f M_f^4}{ M^4 }\right)  \frac{X^2}{L} 
 - {16M_f^3 } \left( \frac{\ddot{M}_f M_f + 2 \dot{M}_f^2}{M^3} \right) \frac{X}{L^2} \label{rhoTG2} \\
& + \frac{16 \dot{M}_f M_f^4 }{M^4} \left( {\dot{M}_f + \int_{V_f}^V \left(3 \ddot{M} + 2 \frac{\dot{M}^2}{M}  \right) d \nu} \right) \frac{X^2}{L^2} 
 \nonumber \\
&  - \left( \frac{16 \dot{M}_f^2 M_1^8}{ M^8 } \right)  \frac{X^4}{L^2} \; . 
 \nonumber
\end{align}

This is still not a simple expression, however, again it is useful. From our experience with the numerical results, we expect any intersections
with the event horizon to occur when $(V_f - V_{int}) L \sim 10$. If this is true then $X_{int}$ will be small and so we should be able to obtain a 
good approximation of $V_{int}$ by finding where the first line of (\ref{rhoTG2}) vanishes. That is we need to solve: 
\bea
  - \frac{4\left(M_f \ddot{M_f} + 2 \dot{M}_f^2 \right)}{M_f \dot{M}_f} 
  = \frac{1}{M(V)} \exp \left( \int_{V_f}^V \frac{L}{4M(\nu)} d \nu \right)  \label{IntEq} \, .
\eea
If $M(V)$ is positive, bounded from below and increasing (which it is) then it is clear that this can only have solutions if $\ddot{M}_f$ is 
negative or (as a stronger condition)
\bea 
\ddot{M}_f \leq -   \frac{2 \dot{M}_f^2}{M_f} \, . 
\eea
Note too that if this bound holds, then there will always be a solution since the right-hand side of equation (\ref{IntEq}) is monotonically 
increasing and goes to zero as $V \rightarrow - \infty$. Equivalently, when the bound is saturated we will necessarily find the intersection value
$V_{int} \rightarrow - \infty$.

Based on our experience in the previous section we expect such a transition to correspond to a peak in the normal-geodesic proper-time length functions. Indeed this seems to be the case. For our particular mass function the bound is saturated when 
\bea
(3+ \mbox{erf} \;  V) V -\frac{2 e^{-V^2} }{\sqrt{\pi}}  = 0 \, , 
\eea
which has solution $V \approx 0.30765$. Examining FIG.~\ref{Lengths} shows that indeed for $L=100$ and $L=1000$, the peak in the length 
function does occur at about this point. 

So the pattern seen in the last section continues into this regime: for our mass function, there is a $V_{peak}$ which divides the 
normal geodesics that intersect the event horizon from those that never intersect it. Let us now consider the proper time lengths of those 
geodesics. It will be sufficient to work at lowest order.  First, from the metric (\ref{scaledmetric}) the (square of the) rate of change of proper time along these geodesics is 
\begin{align}
\frac{1}{m_1^2} \left( \frac{d\tau}{dv} \right)^2 \approx \rho_{TG} L^2 - 4 L \left(\dot{M} + \frac{d \rho_{TG}}{dV} \right) \, . 
\end{align}
So, substituting in 
\bea
\rho_{TG} \approx \frac{4 \dot{M}}{L} + \frac{4 \dot{M}_f}{L} \left(\frac{M_f^4}{M^4} \right) X^2 \, , 
\eea
we find 
\bea
\frac{1}{m_1} \frac{d\tau}{dV} \approx 2 \sqrt{\dot{M}_f L} \left( \frac{M_f^2}{M^2} \right) \exp \left( \int_{V_f}^V \frac{L}{4 M(\nu)} d \nu \,  \right) . 
\eea
Now the proper length from the geometric horizon to the event horizon will be bounded by the integral along the geodesic from $V_f$ to $-\infty$
\bea
\frac{\Delta T}{m_1} \leq \int_{V_f}^{-\infty} \left( \frac{d\tau}{dV} \right) dV 
\eea
and in turn we can use the fact that $M \geq 1$ to bound this integral by
\bea
\frac{\Delta T}{m_1} \leq 8 \sqrt{\frac{\dot{M}_f}{L}} M_f^2 \, . 
\eea

Thus, for any non-decreasing positive mass function $m(v) = m_1 M(v/L)$,  
the proper time distance between the horizons goes to zero as $L \rightarrow \infty$.
In this case the event and geometric horizons are geometrically close. 
 
\subsection{More complex spacetimes}
\label{double}
\begin{figure}
\scalebox {.85}{\includegraphics{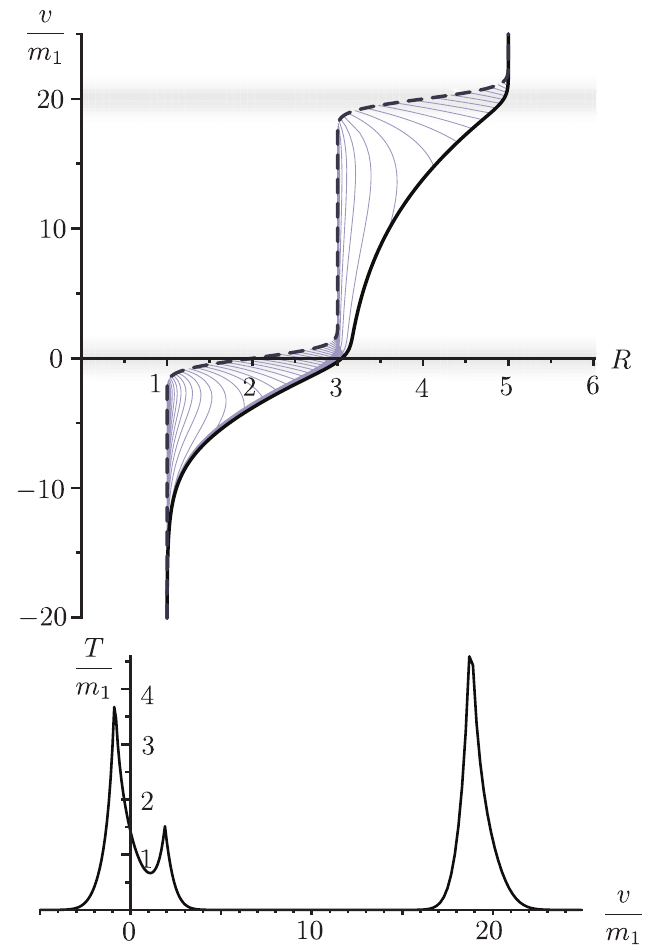}}
\caption{Two shells of dust fall into a black hole causing two distinct phases of expansion. As in the previous figure the top graph shows horizons 
and timelike normal geodesics while the bottom graph is of the proper time between horizons measured along those geodesics. }
\label{HorizonsDouble}
\end{figure}

As a final example we consider a more complicated mass function in which two distinct pulses of dust fall into a pre-existing black hole:
\bea
m(v) = \frac{m_1 (3 + \mbox{erf}(v)+ \mbox{erf}(v-20)) }{2}  \, ,
\eea
The resulting spacetime along with its horizons and geodesics is shown in FIG.~\ref{HorizonsDouble}. 

Turning to the proper time graph, this time there are three peaks and each of these corresponds to a transition between geodesics 
intersecting the event horizon and geodesics which diverge to $- \infty$. The first and third peaks correspond 
to those seen in the earlier examples, while the middle peak arises as some normal geodesics sneak through the gap where the horizons almost
meet and then head off to $v = - \infty$. While the proper time curve continues to record local maximum distances, it is 
clear that this time it does not always record global maxima: the same points of intersection with the event horizon are necessarily 
recorded repeatedly in the lead-up to each of the three peaks. 

Finally note that if we rescaled this mass function as 
\bea
m(v) = \frac{m_1 (3 + \mbox{erf} \left( \frac{v}{L} \right)+ \mbox{erf} \left( \frac{v}{L} -20 \right)) }{2}  \, ,
\eea
and considered the large $L$ limit, all of our near-equilbrium results from the earlier sections would continue to hold. 
 
\section{Discussion}
\label{discuss}

From an astrophysical perspective one can reasonably argue that attempting to define the exact boundary of a black hole is physically about 
as meaningful as discussing the number of angels that can dance on the head of a pin. If one is only interested in what can be observed
from afar then it is the near-horizon fields of the membrane paradigm that are the significant quantities. By construction the horizons are all
unobservable. While this is certainly true in principle, in practical calculations this apparent irrelevance often evaporates: both numerically
and perturbatively horizons of all types are often observed reacting to their environment in intuitive ways. The reason for this is that while 
some particularly dramatic events (such as the actual merger phase of a black hole collision) are far from equilibrium, most black hole physics is 
near-equilibrium. In fact as discussed in \cite{bill} even some very dramatic events, such as black hole formation, can be in this regime.
Then it is reasonable to expect the horizons to be ``close'' together, jointly reflecting the evolution of the external geometry. 


In this paper we have begun to quantify this idea. For the Vaidya spacetime we have located event and apparent horizons and
measured the distance between them using the timelike geodesics normal to apparent horizon. We have seen that at most these proper-times
are of the order of the mass of the black hole while in the near-equilbrium regime time separation goes to zero. If the black hole is eternally 
near-equilibrium the horizons are close together. If there are periodic  bursts of far-from-equilibrium activity, the horizons will separate. However,  
once things begin to calm down they come together again. 

Of course, all of this was done for spherically symmetric surfaces in spherically symmetric spacetimes and those are not real world conditions:
many nice ideas in general relativity have foundered during their generalization from spherical symmetry. That said we believe that a
generalization in this case is quite feasible. It should be possible to perturbatively construct the spacetime geometry around a slowly evolving
(apparent) horizon and we expect that in that geometry it will always be possible to locate an event horizon candidate: a null surface that 
hugs the apparent horizon. Such surfaces have already been located and studied in studies of black brane spacetimes 
\cite{MukundI, MukundII, BoostPaper}. More speculatively one might also hope that this surface will (partially) resolve the non-uniqueness for
apparent horizons. It may turn out that there is a unique null surface close to all near-equilibrium apparent horizons. 
An investigation of these ideas will be the subject of a future paper. 

There is one obvious caveat in our arguments that the exact location of the boundary of a black hole may not matter. For stationary black holes 
it is well-known that the area of the event horizon is directly proportional to the entropy. Further, we have seen that the second-laws of 
horizon mechanics show that horizon area, like entropy, is non-decreasing in time. Thus, it is commonly assumed that entropy continues to be 
proportional to horizon area for dynamical black holes. Then it does become important to pick a single horizon so that entropy will be 
uniquely defined (see \cite{Nielsen} for further discussion on this point). 

Note however that there is another possibility. It is well-known that there is no local definition of gravitational energy 
in general relativity. There may, or may not, be a unique definition of quasilocal energy \cite{QEReview} in stationary spacetimes, however 
far from equilibrium its definition is even more problematic. Now, thermodynamic entropy is operationally defined by the entropy flow
equation $\delta E = T \delta S$. Thus if energy becomes ill-defined or non-unique (not to mention temperature) it would not be surprising 
if there were similar difficulties for entropy. As long as any definition of entropy settles down to the correct value in equilibrium this may be
enough: since all the horizons join (or at least asymptote) together in equilibrium any one of them would be a reasonable candidate. 
Many standard notions of physics (including of course the flow of time) are drastically altered in strong gravitational fields and it is quite
possible that entropy might join the list -- particularly in the regime of non-equlibrium thermodynamics. As long as there are no violations of our
standard laws of physics for observers in the weak-field regime, complications in the strong-field regime are acceptable. Some further discussion of 
these ideas may be found in \cite{BoostPaper} which tries to reconcile notions horizon-defined entropy with those arising from
the AdS-CFT correspondence.  

\acknowledgements{We would like to thank Michal Heller for his useful comments. I.~B. was supported by the Natural Sciences and 
Engineering Council of Canada.}

\begin{appendix}

\section{Measuring infinite geodesics}
\label{appA}
A final application of (\ref{rhoTG}) arises when we wish to find a bound on the total length of a geodesic normal between $V_o$ and $V=-\infty$.
This  situation arises both in cases where the geodesic never intersects the event horizon but instead continues on between the horizons to 
$V=-\infty$ and also in the case where it may intersect for some value of $V < V_o$ but the numerical precision isn't high enough to find
that intersection. In the second case the proper time from $V = - \infty$ to $V_o$ will still provide an upper bound on the true length. 

We proceed as follows. By the metric (\ref{scaledmetric}) the rate of change of proper time along a timelike
curve in the asymptotic regime (where $\rho$ is small and derivatives of $m$ can be neglected) is
\begin{align}
\frac{1}{m_1^2} \left( \frac{d \tau}{dV} \right)^2 \approx  \rho L^2 -  4 L \frac{d \rho}{dV} \, ,
\end{align}
and so by Eq.~(\ref{rhoTG})
\bea
\frac{d \tau}{dV} \approx m_1 L  \sqrt{\rho_o^{TG} -  \frac{4}{L} \left(  \left. \frac{\rho^{TG}}{dV} \right|_{V_o} \right) } \exp \left( \frac{L}{4} (V-V_o) \right)  \, . 
\eea
Then the proper-time length of the timelike geodesic for $V < V_o$ (whether or not it intersects the horizon) is bounded by
\bea
\Delta T = 4 m_1 \sqrt{\rho_o^{TG} -  \frac{4}{L} \dot{\rho}_o^{TG}} \, .
\eea

The values for $\rho_o^{TG}$ and $\dot{\rho}_o^{TG}$ can be calculated from the numerical simulations. For the 
examples considered in FIG.~\ref{Lengths} we find $\Delta T \lesssim 10^{-4} m_1$ is negligible compared to 
the proper time along the geodesic from $V_o$ to $V_f$. 
As such in FIG.~\ref{Lengths}, proper times to the left of the peak are estimated by integrating from $V_o$ to $V_f$.

\end{appendix}

\end{document}